\documentclass[sigplan,screen]{acmart}
\AtBeginDocument{%
  \providecommand\BibTeX{{%
    \normalfont B\kern-0.5em{\scshape i\kern-0.25em b}\kern-0.8em\TeX}}}

\setcopyright{acmcopyright}
\copyrightyear{2022}
\acmYear{2022}
\acmDOI{XXXXXXX.XXXXXXX}
\begin{document}
\acmConference[MSWiM '22]{October 24--28, 2022}{Montreal, Canada}
%
%




%
\title{Efficacy of Asynchronous GPS Spoofing Against High Volume Consumer GNSS Receivers}
\author{M Surendra Kumar}
\email{msurendrakumar2008@gmail.com}
\orcid{1234-5678-9012}
\affiliation{%
  \institution{Department of Aerospace Engineering, Indian Institute of Technology, Bombay}
  \streetaddress{Powai}
  \city{Mumbai}
  \state{Maharashtra}
  \country{India}
  \postcode{400076}
}

\author{Gaurav S Kasbekar}
\affiliation{%
  \institution{Department of Electrical Engineering, Indian Institute of Technology, Bombay}
  \streetaddress{Powai}
  \city{Mumbai}
  \state{Maharashtra}
  \country{India}}
\email{gskasbekar@ee.iitb.ac.in}

\author{Arnab Maity}
\affiliation{%
  \institution{Department of Aerospace Engineering, Indian Institute of Technology, Bombay}
  \streetaddress{Powai}
  \city{Mumbai}
  \state{Maharashtra}
  \country{India}}
\email{arnab@aero.iitb.ac.in}

\renewcommand{\shortauthors}{M Surendra Kumar, et al.}

\begin{abstract}
  The vulnerability of the Global Positioning System (GPS) against spoofing is known for quite some time. Also, the positioning and navigation of most semi-autonomous and autonomous drones are dependent on Global Navigation Satellite System (GNSS) signals. In prior work, simplistic or asynchronous GPS spoofing was found to be a simple, efficient, and effective cyber attack against L1 GPS or GNSS dependent commercial drones. In this paper, first we make some important observations on asynchronous GPS spoofing attacks on drones presented in prior research literature. Then, we design an asynchronous GPS spoofing attack plan. Next, we test the effectiveness of this attack against GNSS receivers (high volume consumer devices based on Android mobile phones) of different capabilities and a commercial drone (DJI Mavic 2 Pro) under various conditions. Finally, we present several novel insights based on the results of the tests.
\end{abstract}



\maketitle

\section{Introduction}
Unmanned Aerial Vehicles (UAVs) or drones are no more a technology of the future; they are a current story unfolding \cite{article}. Nearly all the businesses, activities, and services carried out by humankind on this planet are on the verge of being disrupted by this technology \cite{boyle2020drone}. Indian national planners too have taken cognizance of this possibility and have liberalized the civilian drone rules along with introducing the Production Linked Incentives (PLI) in our country \cite{kumar2021sky}. The impact of these changes made is seen as a catalyst in expanding the Indian drone industry, which is presently having a turnover of approximately Indian National Rupees (INR) 80.00 crores, to an estimated INR 900.00 crores by 2023 and INR 15,000.00 crores by 2026 \cite{karunakaran2022swarm}. It is also estimated that every year, 10 million plus GNSS enabled drones will be manufactured and shipped this decade \cite{EUSPAEO}. This exponential growth in the number of drones around us makes it imperative for us to assess the vulnerabilities of commercially available UAVs against various threats and be prepared to exploit or counter them. 
\begin{figure}[h]
  \centering
  \includegraphics[width=\linewidth]{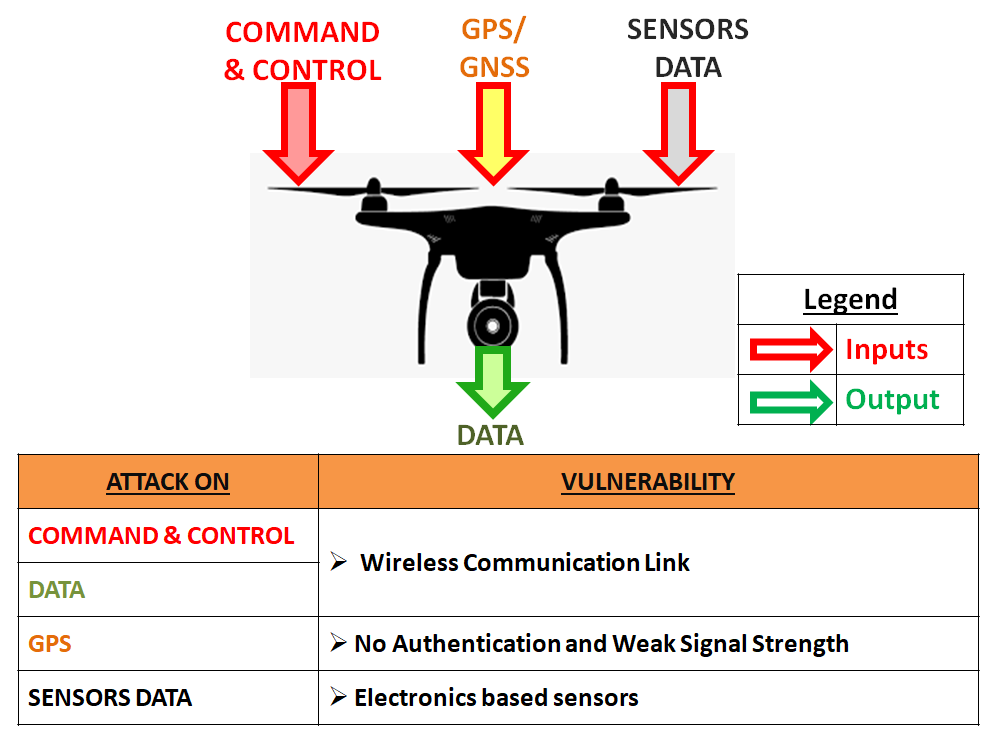}
  \caption{Ingress and egress routes, i.e., inputs into and outputs from a drone along with their vulnerabilities.}
  \Description{Inputs and outputs from a typical semi autonomous or autonomous drone along with related vulnerabilities.}
  \label{IERoutes}
\end{figure}

Researchers worldwide have studied cyber-security vulnerabilities of UAVs \cite{hartmann2013vulnerability, krishna2017review, yaacoub2020security, gordon2019security, tufekci2021vulnerability}. Cyber threats are often considered equivalent to theft or disruption, and these are carried out using the ingress or egress routes \cite{hathaway2014melissa}.  Fig. \ref{IERoutes} shows the vulnerabilities in ingress and egress routes, i.e., the inputs to and outputs from a UAV. In semi-autonomous and autonomous drones, the GPS or GNSS input is vital since in addition to basic positioning and reporting locations on maps, other functions such as return to home and no-fly zone are based on them. From previous work done in this field, it was found that GPS spoofing is an effective attack and it is applicable against a large number of commercial drones.   

Based on the success of GPS spoofing against commercial drones in previous research works, our main aim was to check if GPS spoofing can be used as an effective countermeasure against rogue commercial drones. As more than 90\%  of high volume consumer devices support only the L1 frequency \cite{EUSPA}, in this work we set up an asynchronous GPS spoofing attack and test its efficacy against GNSS receivers (single frequency multi-constellation based receivers with different capabilities in high volume consumer devices such as Android mobiles) and a commercial drone.  The major contributions of this paper are as follows: \begin{itemize}
    \item We first observe the commonality in the conditions under which GPS spoofing attacks reported in prior research literature were conducted.
    \item Next, we design a detailed GPS spoofing attack plan.  
    \item We generate GPS spoofing signals and transmit them.
    \item We test our attacks against GNSS receivers and a commercial drone under different conditions.
    \item We present our observations from the results.
\end{itemize}

The rest of this paper is organized as follows. Section 2 provides a review of prior research literature. We  explain our GPS spoofing attack plan in Section 3. Section 4 describes the experimental setup and scenarios. In Section 5, the results obtained from our experiments are presented and discussed. Section 6 concludes our work. 

\section{Related Work}
	\label{lit}
	Commercial drones can be categorized into three types, based on the role of an operator and its autonomy \cite{aabid2022reviews}: 1) Remotely Piloted Vehicle: Driven by drone pilots, always keeping the drone within their visual range, 2) Semi-Autonomous: Has capabilities such as hovering and returning to home without inputs by the operator, 3) Fully-Autonomous: Does not necessarily require a Command-and-Control (C2) link. 
	

\begin{table*}
  \caption{System and conditions of GPS spoofing}
  \label{tab:table3}
  \begin{tabular}{ccl}
    \toprule
    Reference & System & Conditions\\
    \midrule
    {\cite{horton2018development}} & SDR (BladeRF X40)$+$Open & •	Static and dynamic spoofing done against static drone \\ 
&  Source Simulator (OSS) & •	Experiment carried out indoor\\
\hline
{\cite{he2018friendly}} & SDR (USRP B210)$+$ & •	Target's position required \\ 
& Replayed received signal & •	Target drone in loiter or hovering mode\\
\hline
{\cite{dey2018security}} & LABSAT3 & •	Experiment carried out indoor \\ 
& GPS simulator & •	Drone in static position\\
\hline
{\cite{arteaga2019analysis}} & SDR (BladeRF X40) & •	Experiment carried out indoor \\ 
& $+$OSS & •	Drone landed or hovering\\
\hline
{\cite{zheng2020hijacking}} & SDR (HackRF One)$+$OSS & •	Attacks carried out against hovering or landed drone \\ 
\hline
{\cite{sathyamoorthy2020evaluation}} & Aeroflex GPSG & •	Experiment carried out under lab conditions \\ 
& -1000 GPS simulator & •	In absence of live-sky signals\\
\hline
{\cite{saputro2020implementation}} & SDR (BladeRF X40)$+$OSS & •	Attacks carried out against hovering drone \\
    \bottomrule
  \end{tabular}
\end{table*}
    The GPS is an all-weather, satellite-based navigation system developed by the Department of Defence (DOD) of the United States. It was developed for military use, but today it continuously provides position, velocity, and time (PVT) services to military as well as civilian users worldwide at or near the surface of the earth  \cite{mcneff2002global,GPSSignals}. Satellite navigation systems with global coverage are termed GNSS \cite{blanch2012satellite}. E.g., United States’ GPS, Russia’s GLONASS, China’s BeiDou navigation satellite system, and the European Union’s Galileo are the four operational GNSS. Japan and India have their regional satellite-based navigation systems \cite{blanch2012satellite}. GPS being the earliest in becoming fully functional with global coverage and free of cost, has been widely used commercially. 100\% of GNSS receivers manufactured till date have L1 frequency compatibility \cite{lisi2020gnss}. Moreover, over 70\% of the GNSS receivers can operate only on the L1 frequency \cite{lisi2020gnss}.
    
The vulnerability of L1 GPS signals is known for almost two decades now. Civilian GPS signals are vulnerable due to three reasons:
\begin{itemize}
    \item  Transparency of the technology\cite{mcneff2002global,GPSSignals,blanch2012satellite,misraglobal}: Owing to this, developers have developed software to generate signals similar to GPS. An open source GPS simulator is \cite{GPS-SDR-SIM} (commercially too GPS or GNSS simulators \cite{broadsim,satgen} are available).
    \item The fact that there is no authentication mechanism in civilian L1 GPS signals keeps it vulnerable to spoofing \cite{mcneff2002global,GPSSignals,blanch2012satellite,misraglobal}.
    \item The low power of the GPS received signals \cite{mcneff2002global,GPSSignals,blanch2012satellite,misraglobal} keeps it vulnerable to being over-powered.
\end{itemize}

The GPS spoofing threat was assessed and various versions of GPS spoofing were explained by the authors in \cite{humphreys2008assessing}. Different versions of GPS spoofing include asynchronous spoofing (when an attacker does not synchronize the spoofed signals with the original GPS signals), intermediate spoofing (in this, an attacker  synchronizes the spoofed signal based on the knowledge of the receiver’s location and time), sophisticated spoofing (in this, an attacker transmits synchronized spoofed signals using multiple antennae to overcome the direction of arrival counter-measure), and meaconing (in this spoofing methodology, an attacker re-transmits the received authentic signals) \cite{humphreys2008assessing}. The authors of  \cite{humphreys2008assessing} also developed a sophisticated GPS spoofer, but it was too bulky and costly.  The authors of \cite{tippenhauer2011requirements} explained the nuances of carrying out successful GPS spoofing attacks; however, most of the experiments in this paper were set up in a controlled environment. In \cite{shepard2012drone}, the authors demonstrated hacking a commercial civilian UAV using GPS spoofing. The hacking was carried out under the live sky, but used specifically designed GPS spoofing equipment. GPS spoofing using a Software Defined Radio (SDR) was demonstrated in \cite{huang2015low}. A software developer developed code for a GPS simulator and made it public by uploading the same on GitHub \cite{GPS-SDR-SIM}. This simulator generated L1 civilian GPS baseband signal data streams and required an SDR for generation and transmission (Tx) of Radio Frequency (RF) signals. In \cite{huang2015low} and \cite{GPS-SDR-SIM},  the equipment and process for carrying out an asynchronous GPS spoofing attack were simplified. Numerous research works have been done since then, in the area of GPS spoofing simulations and attacks, showing the vulnerabilities of various commercial drones using open source code or commercial GPS simulators \cite{horton2018development,he2018friendly,dey2018security,arteaga2019analysis,zheng2020hijacking,sathyamoorthy2020evaluation,noh2019tractor,saputro2020implementation}. Nearly all of the GPS spoofing work done till date have been reviewed in \cite{khan2021gps}. The authors did not limit their review to GPS spoofing attacks against aerial platforms, but also included non aerial platforms. 

Table \ref{tab:table3} tabulates some of the recent successful GPS spoofing attacks against commercial civilian drones along with the spoofing system utilised and the specific conditions under which attacks were carried out.
It can be observed from Table \ref{tab:table3} that most of the spoofing attacks have been carried out against static or hovering drones and indoors (in absence of live sky signals). Moreover, all the attacks were against GPS receivers and not against GNSS receivers.   In \cite{zidan2020gnss}, the authors summarized various countermeasures against GNSS spoofing, along with the sophistication and cost required in implementing them.  Table \ref{tab:table3} shows that under the conditions listed in the table, GPS spoofing attacks are successful against commercial drones. However, in \cite{rustamov2020assessment}, it has been shown that mass consumer devices like Android mobiles with dual-frequency band GNSS chipset are resilient against synchronous GPS spoofing attacks. In this paper, we seek to evaluate the efficacy of asynchronous GPS spoofing, find out the conditions under which it works and hence can be used as an effective countermeasure against rogue commercial drones, and to suggest countermeasures to protect drones against it.
 
\section{GPS Spoofing Attack Plan}
\label{model}
\begin{figure}[h]
  \centering
  \includegraphics[width=\linewidth]{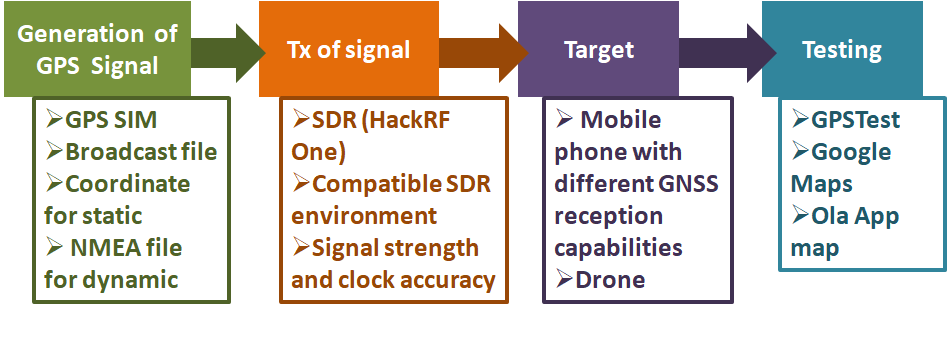}
  \caption{GPS spoofing attack plan designed considering generation, transmission (Tx), targets of attack, and method to test the results.}
  \Description{Designed and executed GPS spoofing attack plan consisting of various phases like generation, Tx, testing, and targets.}
  \label{Atkp}
\end{figure}

 The GPS spoofing attack plan was designed and executed to check the efficacy of asynchronous GPS spoofing against GNSS receivers. The attack plan consists of the following phases: generation, Tx, target, and testing. Fig. \ref{Atkp} shows our attack plan. 

\subsection{Generation of GPS Spoofed Signal}
\begin{figure}[h]
  \centering
  \includegraphics[width=\linewidth]{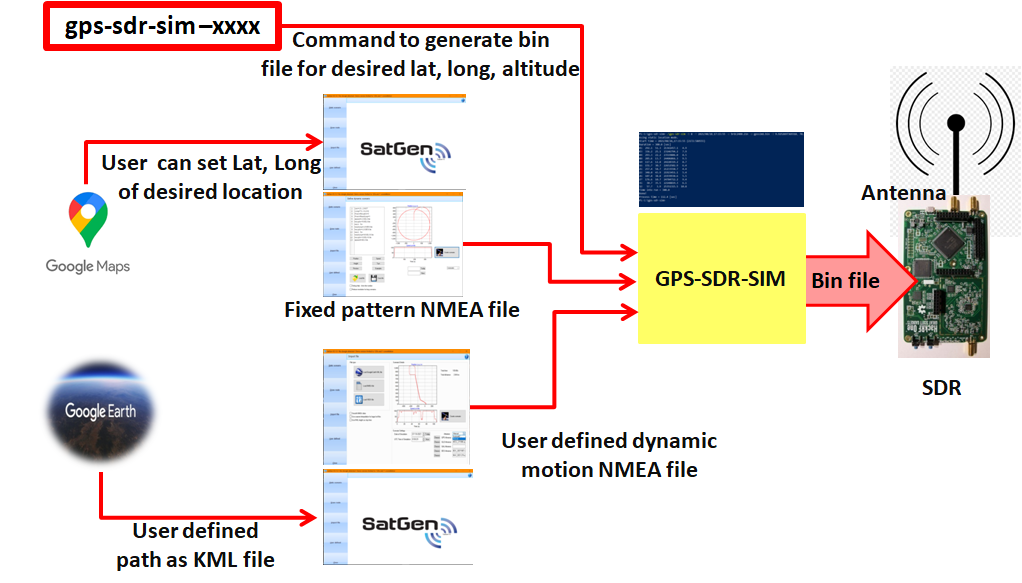}
  \caption{The image shows how generation of various spoofed GPS signals (static or dynamic) has been carried out.}
  \Description{Pictorial representation of the process used for generation of spoofed GPS signal.}
  \label{genr}
\end{figure}

The generation of a GPS baseband signal has been carried out on a laptop with Windows 10 as its operating system.  GPS-SDR-SIM \cite{GPS-SDR-SIM} is used as the GPS simulator in our experiments. GPS-SDR-SIM is an open source GPS simulator written in C, which can generate GPS baseband data streams in binary format. A Receiver INdependent EXchange format (RINEX) navigation file for GPS ephemerides is required by GPS-SDR-SIM, which can be downloaded from the cddis.nasa website \cite{Brdc}. The broadcast ephemeris files are named as brdcDDD0.YYn, where DDD is the number of days of the year, i.e., 001 to 365/ 366 and YY are the last two digits of the year. E.g., for 10 Jun 2022, the broadcast ephemeris file is uploaded with the name brdc1610.22n. The broadcast ephemeris file data is uploaded post 4 hours; thus, spoofing was not carried out for real-time scenarios. For static spoofing (i.e., when the generated spoofed signals make a GPS receiver compute its position to be a static location), GPS-SDR-SIM requires the coordinates of the destination in either x,y,z in the Earth Centered Earth Fixed (ECEF) coordinate system or latitude (lat), longitude (long), and altitude (alt) in the World Geodetic System (WGS 84 coordinate system). We used the lat, long, and alt data and sourced these information of the desired spoofed destination locations from Google Maps. For dynamic spoofing (i.e., when the generated spoofed signals make a target GPS receiver compute its position to follow a dynamic path with a certain velocity),  GPS-SDR-SIM requires either a user defined Comma Separated Values (CSV) file or a National Marine Engineering Association (NMEA) GGA stream of data. We utilised Google Earth to define a route at the desired location, exported the Keyhole Markup Language (KML) file onto the SatGen software \cite{satgen} (Labsat provides a trial version of its SatGen software which can generate a  L1 Coarse Acquisition (C/ A) signal) and generated the NMEA GGA data stream. Fig. \ref{genr} is a pictorial representation of how the spoofed GPS signal is generated.
\subsection{Tx of Generated Spoofed Signal}
The generated GPS baseband data streams are converted to RF using a Software Defined Radio (SDR). We have utilized the HackRF One SDR (the lowest price of HackRF One SDR in comparison to other available SDRs was the reason for choosing it). For Tx of the generated RF signal, a compatible 700-5800 MHz  directional patch array antenna was utilised. The intention behind using a directional antenna was to keep the generated spoofed signals in the desired directions only, while carrying out the experiments. The internal crystal clock accuracy of our SDR is 20 parts per million (PPM). To improve the accuracy, we used a Temperature Compensated Crystal Oscillator (TCXO) with accuracy 0.5 PPM. Fig. \ref{exptsu} (top left) shows a TCXO connected to a HackRF One SDR.

\subsection{Target and Testing}
\begin{figure}[h]
  \centering
  \includegraphics[width=\linewidth]{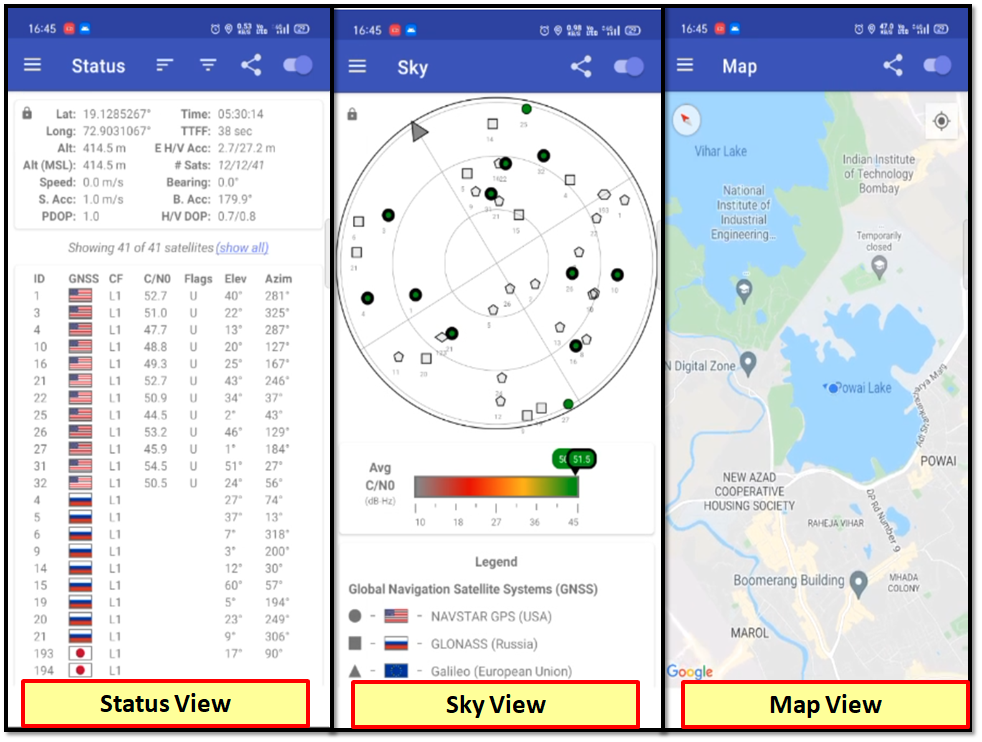}
  \caption{GPSTest App's various views: status view (left), sky view (centre), and map view (right).}
  \Description{Figure showing screenshots taken on Android device while running GPSTest App with various views of it.}
  \label{GTV}
\end{figure}

The efficacy of an asynchronous GPS spoofing attack was tested against GNSS receivers in high volume consumer devices with different chipsets and support for different constellations as shown in Table \ref{tab:table4}, and against a commercial drone (DJI Mavic 2 Pro). To record the received signals in the Android devices, we utilised the GPSTest Application (App)  \cite{GPSTest}. Fig. \ref{GTV} shows various views of the GPSTest App. The leftmost screenshot is of the status view. The status view gives  a lot of information; the  important ones for understanding our results are:
\begin{itemize}
    \item Lat and Long, both in decimal degrees WGS 84, Alt (meters above the WGS 84 ellipsoid)
    \item Time To First Fix (TTFF), i.e., the time taken by the GNSS receiver to compute its position in seconds
    \item Speed in m/ s
    \item Number of satellites (\#Sats) in the format xx/ yy/ zz: where xx shows the number of satellites used in the fix, yy represents the number of satellites in view with valid signal strength, and zz shows the number of satellites known to the device
    \item C/N0: Carrier to noise density ratio in dBHz.
\end{itemize}
The centre screenshot is of the sky view and the rightmost screenshot is of the map view. In addition to the GPSTest App's map, we also utilised Google Maps and Ola App's map for visualising our original and spoofed locations on the Android devices. DJI Go 4 App is the requisite App required by a DJI mobile based controller to fly a DJI drone. We utilised this App to visualize the drone's location, registered the drone's home location, and controller location. Our testing was conducted in three phases. 
\begin{itemize}
    \item Phase I: Testing different scenarios onto a single Android mobile device.
    \item Phase II: Testing different scenarios onto all four Android mobile devices simultaneously.
    \item Phase III: Testing GPS spoofing attack on drone.
\end{itemize}

\begin{table*}
  \caption{Models of Android Mobiles, Chipsets and GNSS Capabilities}
  \label{tab:table4}
  \begin{tabular}{ccccc}
    \toprule
   \textbf{Sr. No.}$^{\mathrm{@}}$&\textbf{Build}&\textbf{Model}&\textbf{Chipset}&\textbf{GNSS Capability} \\
    \midrule
    S1&	Realme	3&  RMX1825&	Mediatek Helio P70& G+L+A \\
    S2&	Oppo&	F7& Mediatek MT6771(Helio P60)& G+L+A \\
    S3&	ASUS&	Max Pro M1& Qualcomm SDM 636 Snapdragon&	G+L+B+A \\
    S4&	POCO&	M2&	Mediatek MT6769V/CU Helio G80&	G+L+B+A\\
    \hline
    \multicolumn{5}{c}{\textbf{G= GPS, L= GLONASS, B= BeiDou, A= Assisted GPS}} \\
    \bottomrule
\multicolumn{5}{p{.9\textwidth}}{$^{\mathrm{@}}$While assessing the results, we will be referring to different Android devices with their sr. nos., i.e., S1, S2, S3 and S4.}
      \end{tabular}
\end{table*}

\section{Experimental Scenarios and Setup}

\subsection{Experimental Scenarios}
\begin{figure}[h]
  \centering
  \includegraphics[width=\linewidth]{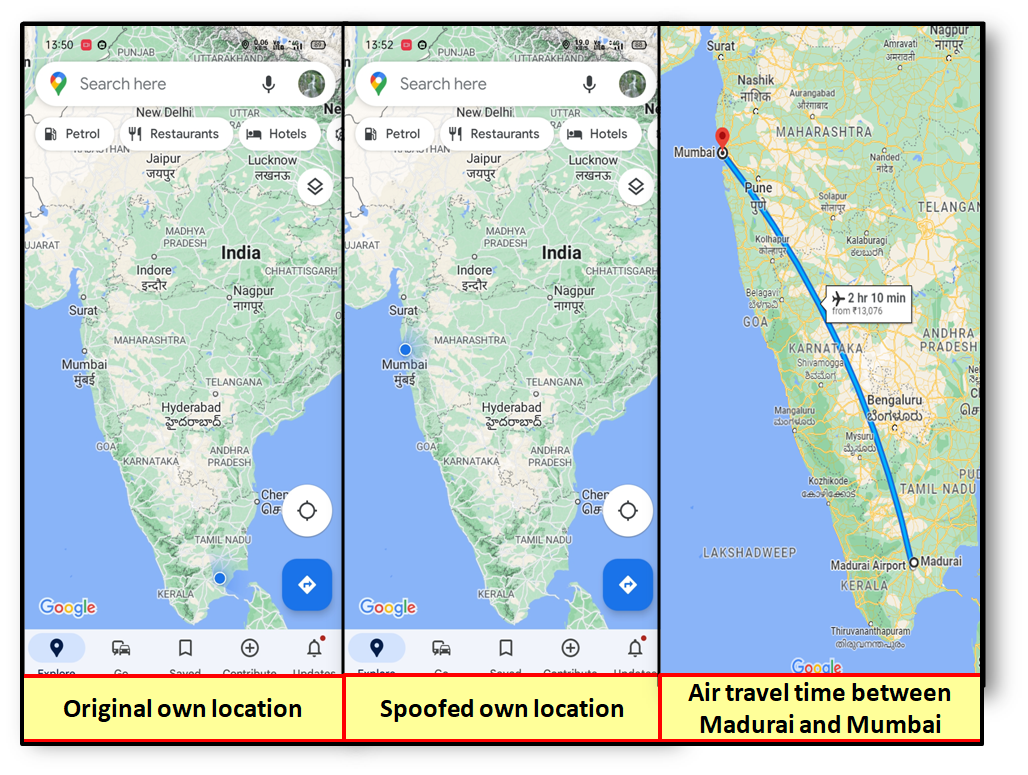}
  \caption{Screenshots of original own location (left), spoofed own location (center), air travel duration between the two (right).}
  \Description{Figure showing screenshots taken on Android device location of original own location, spoofed own location and flight duration between the two locations.}
  \label{oso}
\end{figure}

The efficacy of an asynchronous GPS spoofing attack against a GNSS receiver in an Android mobile phone was tested under the following conditions: 
\begin{enumerate}
    \item Receiver placed indoor/ outdoor in the absence/ presence of live sky signals, respectively.
    \item Static/ dynamic spoofed signal is transmitted.
    \item Receiver is static/ mobile.
\end{enumerate}
The simulated GPS signals were designed to simulate static and dynamic locations in and around Indian Institute of Technology, Bombay (Mumbai), Maharashtra, India. The experiments were conducted in the outskirts of Madurai, Tamil Nadu, India. The physical crow fly distance between the two locations is nearly 1200 kms. Fig. \ref{oso} shows the geographical location of the original own location (left), spoofed own location (center), and the air travel duration between the two locations, which is more than 2 hrs and 30 mins (right).
\subsection{Experimental Setup}
\begin{figure}[h]
  \centering
  \includegraphics[width=\linewidth]{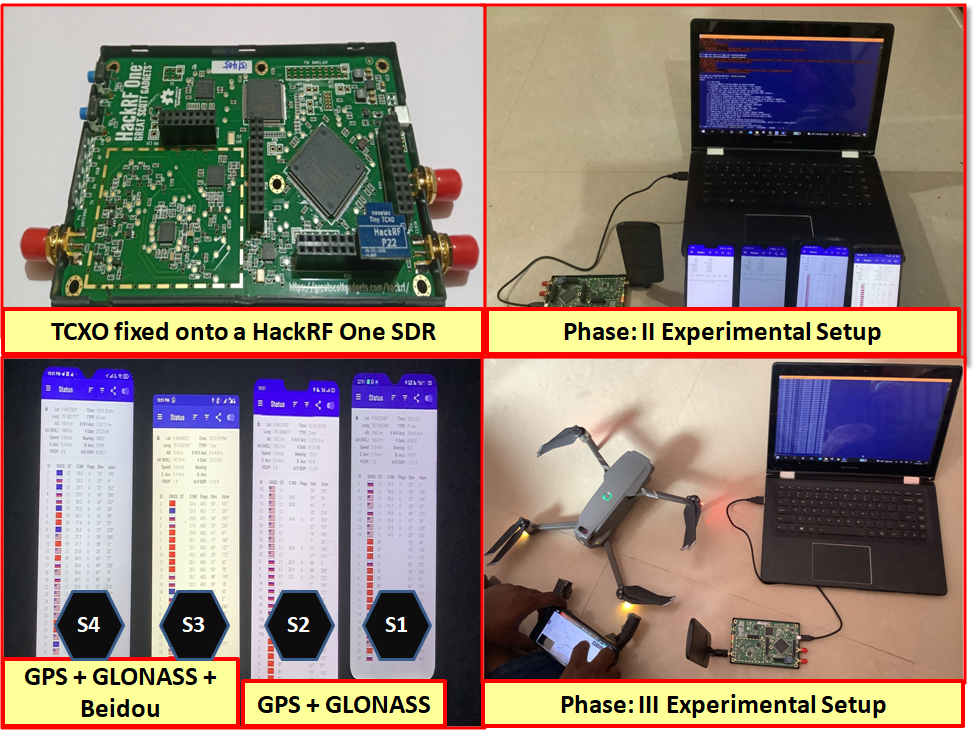}
  \caption{Top left: external TCXO connected on HackRF One SDR; Top right: indoor experimental setup; Bottom left: capabilities of different GNSS receivers; Bottom right: setup for spoofing test against drone.}
  \Description{Top left: Image showing external TCXO fixed onto the HackRF One SDR. Top right: Image showing indoor experiment setup for testing multiple GNSS receivers of different capabilities. Bottom left: Image showing different GNSS capabilities of tested Android mobile phones. Bottom right: Image showing setup for carrying out GPS spoofing test on a commercial drone i.e. DJI Mavic 2 Pro.}
  \label{exptsu}
\end{figure}
The  top (right) and bottom (right) images in Fig. \ref{exptsu} show the experimental setup for conducting the efficacy test against Android mobile phone based GNSS receivers and a commercial drone (DJI Mavic 2 Pro), respectively. As per the user manual, a DJI Mavic 2 Pro's GNSS receiver supports GPS and GLONASS. However, the frequencies and bands utilised by the drone are not shared. DJI drones use a DJI GO App, which can be run on an Android or iOS mobile phone and used as a controller of the drone. We used both Android and iOS mobile phones as the controller and our results in the two cases were the same (see Section \ref{Expt}). 

\section{Experimental Results and Discussion}
\label{Expt}
Initially, the spoofed signals transmitted by the SDR as plan- ned in Fig. \ref{Atkp} were not received by the Android phone GNSS receiver. However, once an external TCXO with 0.5 PPM accuracy was fixed onto the HackRF One SDR as shown in Fig. \ref{exptsu} (top left), the spoofed signals were received by the GNSS receiver. Figs. \ref{ISS}-\ref{DTOios} (explained in the following sections) are  screenshots of the GPSTest App, Google Maps, OLA App, camera capture, DJI Go 4 map, and Apple map showing the results of spoofing taken during testing. Table \ref{tab:table5} shows the results of various asynchronous GPS spoofing tests conducted over the embedded GNSS receiver of an Android mobile phone S1 (see Table \ref{tab:table4}).

\subsection{Indoor and Outdoor (Static/ Dynamic) Spoofing Against an Android Mobile Phone based GNSS Receiver}
\begin{figure}[h]
  \centering
  \includegraphics[width=\linewidth]{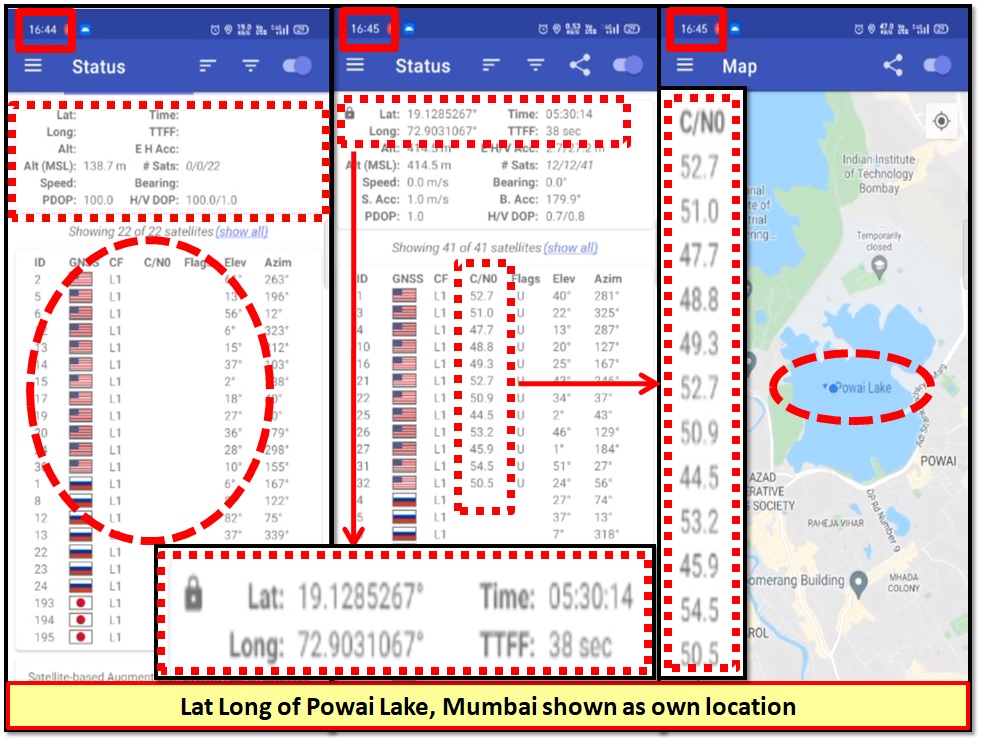}
  \caption{Indoor static spoofing: the leftmost screenshot is of GPSTest App with no original GNSS signal being received, center screenshot shows the reception of the spoofed signal and the GNSS receiver locking onto it, and the rightmost screenshot of Google Maps shows own location as in the middle of Powai lake, Mumbai.}
  \Description{Figure showing screenshots taken on Android device while carrying out indoor static spoofing.}
  \label{ISS}
\end{figure}
Fig. \ref{ISS} shows the screenshots (left to right) taken during the static GPS spoofing attack carried out against a GNSS receiver under indoor conditions. The GNSS receiver was not able to receive any signals from GPS or GLONASS constellation satellites, as highlighted in the leftmost screenshot. Once the spoofed signals were transmitted, the GNSS receiver started receiving them instantaneously and was able to compute its position within 30-40 s. The center screenshot highlights that it took the receiver 38 s as TTFF.   The received C/N0 is found to be ranging from 40 to 55 dBHz.
\begin{figure}[h]
  \centering
  \includegraphics[width=\linewidth]{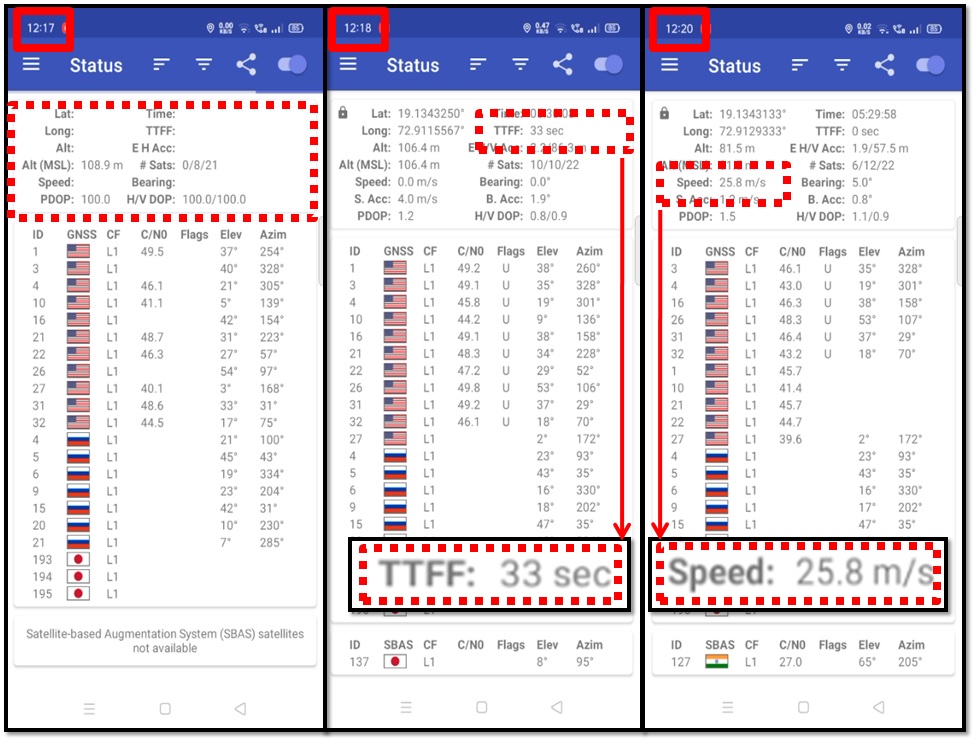}
  \caption{Indoor dynamic spoofing: screenshots of GPSTest App. The leftmost screenshot shows the beginning of spoofing. The middle screenshot shows the TTFF and the rightmost screenshot shows speed.}
  \Description{Figure showing screenshots taken on Android device while carrying out indoor dynamic spoofing.}
  \label{IDS}
\end{figure}
Fig. \ref{IDS} shows the effect of indoor dynamic spoofing on the GNSS receiver. As was seen in Fig. \ref{ISS} itself, the receiver was not able to receive any genuine signals indoors. Fig. \ref{IDS} shows screenshots from the beginning of the dynamic spoofing till the receiver computed its position (left to right). Even though the TTFF is shown as 33 s, the own location blip in Google Maps took some time (3-10 s) to follow the dynamic movement. The rightmost screenshot of Fig. \ref{IDS} highlights the receiver showing its speed as 25.8 m/ s.

\begin{figure}[h]
  \centering
  \includegraphics[width=\linewidth]{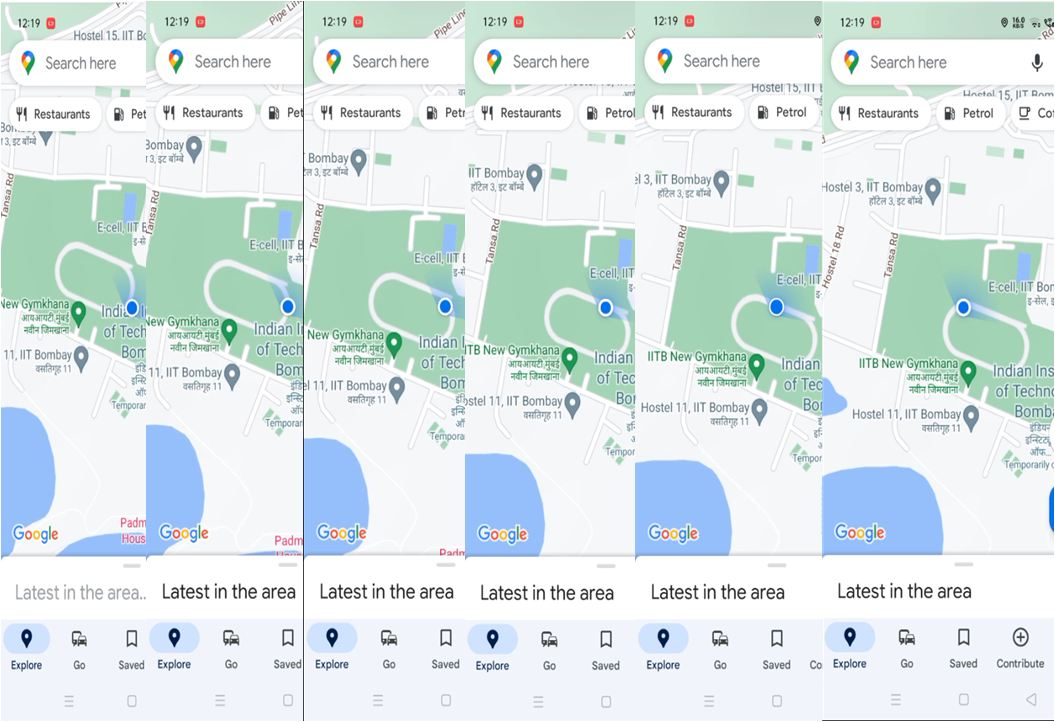}
  \caption{Indoor dynamic spoofing: screenshots of Google Maps App. Left to right are the screenshots showing the spoofed location as if moving around the Gymkhana ground track in IIT Bombay.}
  \Description{Figure showing screenshots of Google Maps app running, taken on Android device while carrying out indoor dynamic spoofing.}
  \label{IDS22}
\end{figure}

\begin{figure}[h]
  \centering
  \includegraphics[width=\linewidth]{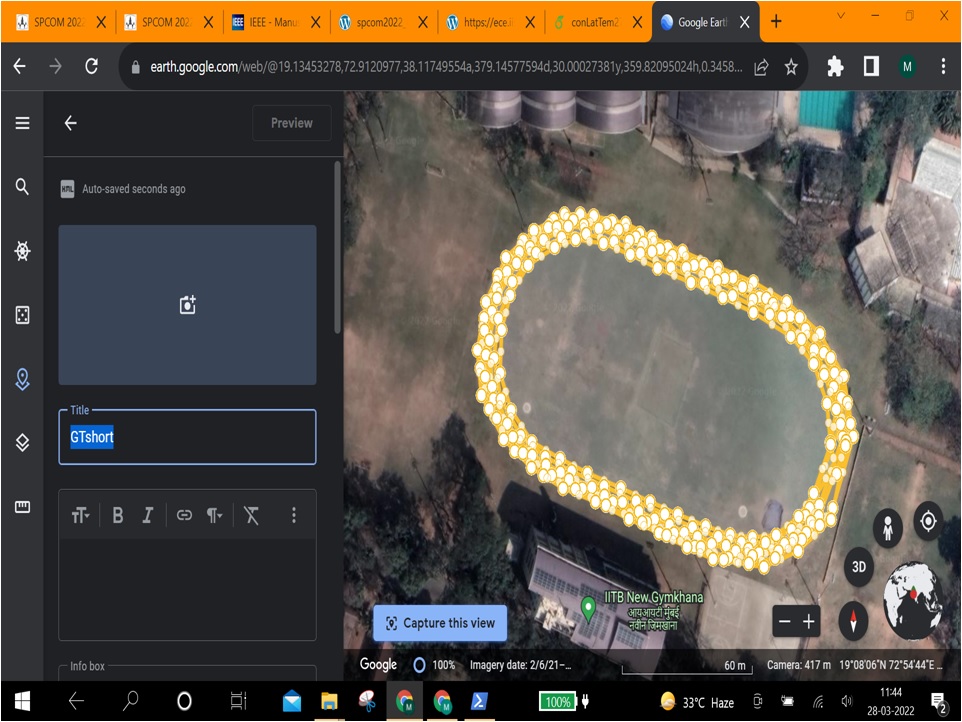}
  \caption{Designing route in Google Earth with numerous waypoints.}
  \Description{Screenshot of Google Earth taken while the route was drawn which will be subsequently used for generation of dynamic spoofing file.}
  \label{designwp}
\end{figure}

Fig. \ref{IDS22} shows sequential screenshots (left to right) of the Google Maps App running and showing its location to be shifting around the Gymkhana track of IIT Bombay (as designed in the spoofed signal generation process). The initial results of dynamic spoofing were not consistent, i.e., the movement of own location blip in Google Maps was not corresponding accurately with the designed route. To improve the same, the way-points, while designing the route in Google Earth, were selected at equal distances with several repetitions (see Fig. \ref{designwp}); this improved the spoofing results.
\begin{table*}
  \caption{Results of Phase I Tests}
  \label{tab:table5}
  \begin{tabular}{c|cc|cc}
    \toprule
   \textbf{Location of Receiver}&\multicolumn{2}{c|}{\textbf{Indoor}}&\multicolumn{2}{c}{\textbf{Outdoor}}\\
  \hline
\textbf{Receiver}&\textbf{Static}&\textbf{Mobile*}&\textbf{Static}&\textbf{Mobile*} \\
    \midrule
    \textbf{\textcolor{blue}{Static spoofed signal}}& \textcolor{blue}{S}(30-40 s)&	\textcolor{blue}{S}(30-40 s)&	\textcolor{blue}{S}(30-50 s)&	\textcolor{blue}{S}(>60 s)\\
\textbf{\textcolor{blue}{Dynamic spoofed signal}}& 	\textcolor{blue}{S}(30-40 s)&	\textcolor{blue}{S}(>60 s)&	\textcolor{blue}{S}(>90 s)&	\textcolor{blue}{S}
(>90 s)\\
   \hline
\multicolumn{5}{c}{\textcolor{blue}{S}= spoofing was successful; (xx s) = time taken by the receiver to show spoofed location.}\\
    \bottomrule
\multicolumn{5}{p{.95\textwidth}}{$^{\mathrm{*}}$ Movement to the receiver was imparted by manual walking movement with a speed not more than 1 meter (m) per second (s). Also, the maximum distance of the receiver from the spoofer during the experiments was less than 50 m. All the tests  were carried out with the cellular 4G network on.}
     \end{tabular}
\end{table*}

\begin{figure}[h]
  \centering
  \includegraphics[width=\linewidth]{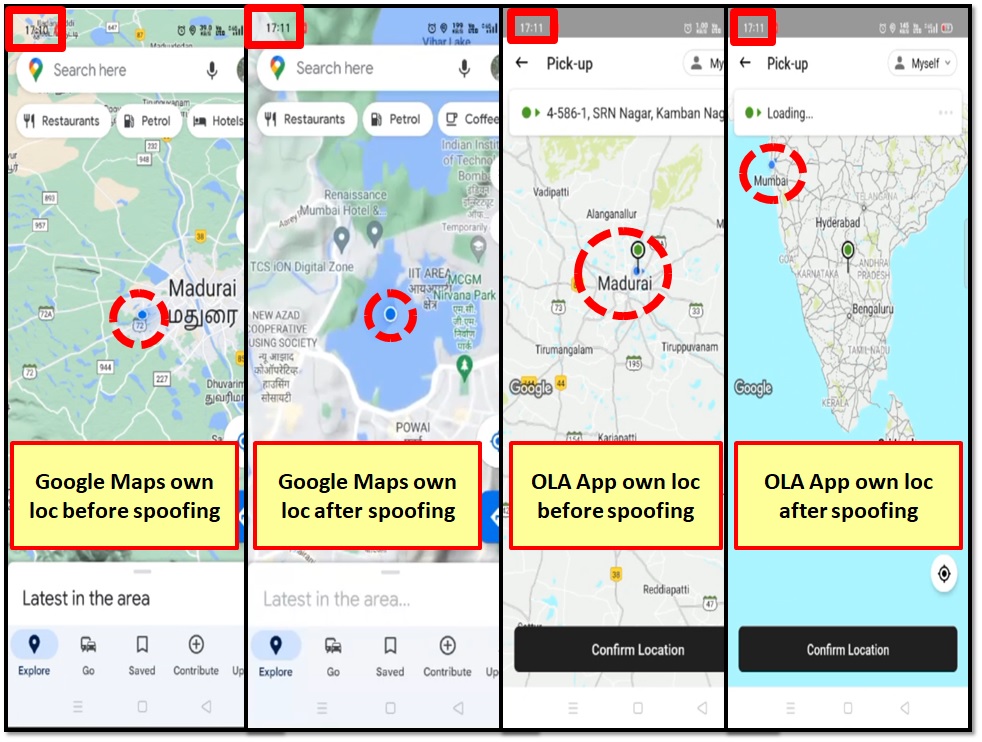}
  \caption{Outdoor static spoofing: the two screenshots on the left  are of the Google Maps App before and after spoofing. The two screenshots on the right are of the OLA App map before and after spoofing.}
  \Description{Results of outdoor static spoofing shown by means of screenshots of Android device taken with Google Maps App and Ola App Map running with and without spoofed signals.}
  \label{24}
\end{figure}

\begin{figure}[h]
  \centering
  \includegraphics[width=\linewidth]{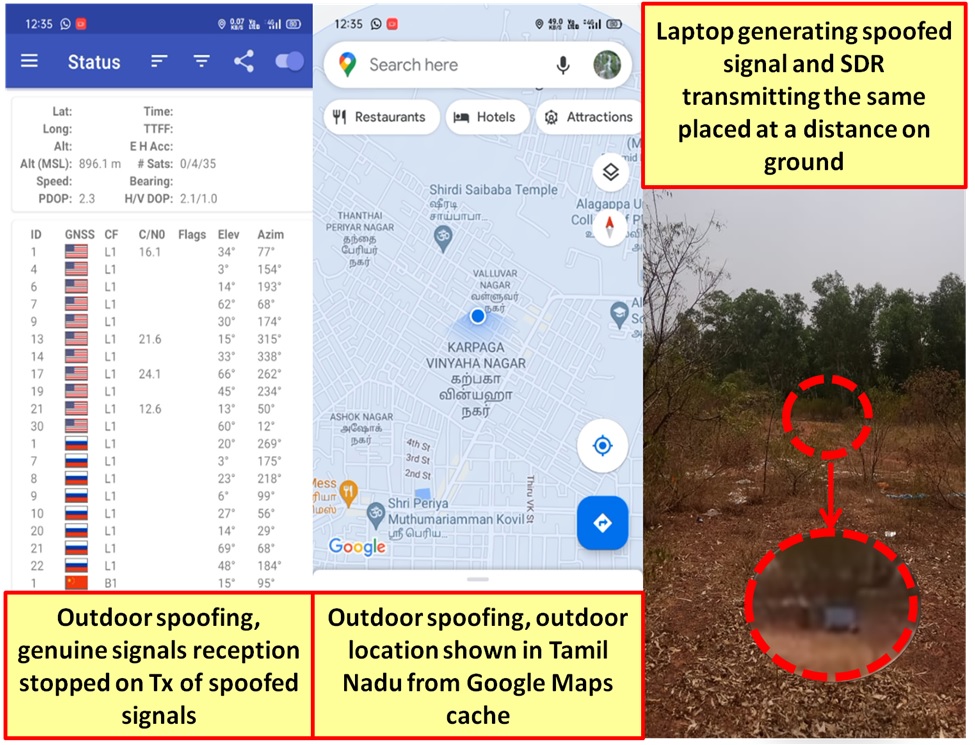}
  \caption{Left to right: screenshots of GPSTest App, Google Maps, and mobile phone camera during outdoor dynamic spoofing.}
  \Description{Results of outdoor dynamic spoofing: shown by means of screenshots of Android device taken with GPSTest App, Google Maps, and mobile phone camera before receipt of spoofed signals.}
  \label{25}
\end{figure}

\begin{figure}[h]
  \centering
  \includegraphics[width=\linewidth]{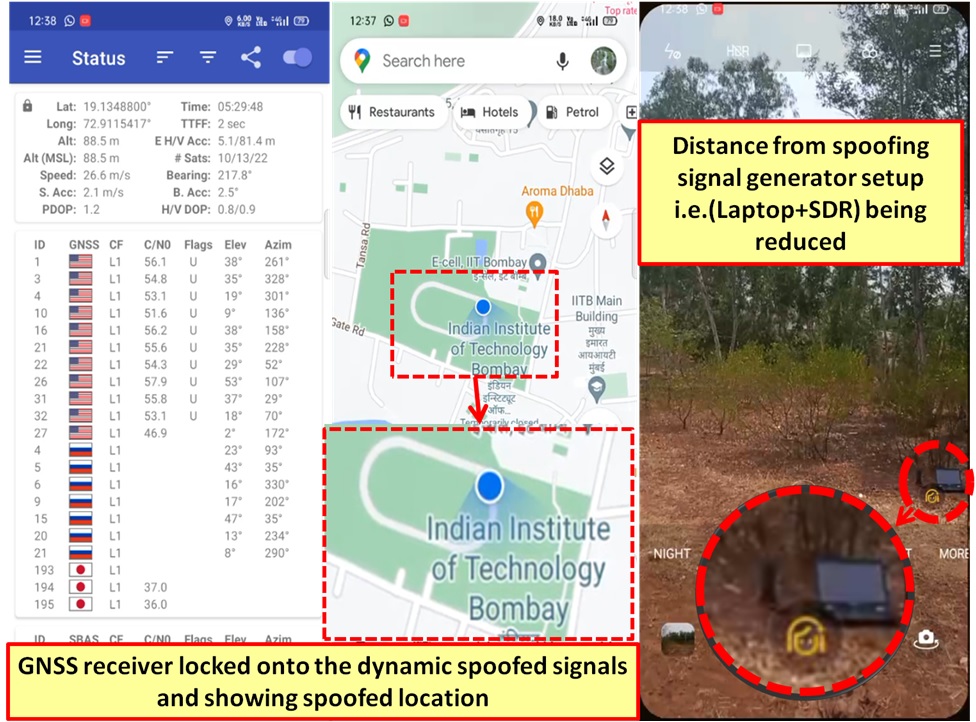}
  \caption{Left to right: screenshots of GPSTest App, Google Maps, and mobile phone camera after receipt of dynamic spoofing signals (outdoor).}
  \Description{Results of outdoor dynamic spoofing: shown by means of screenshots of Android device taken with GPSTest App, Google Maps, and mobile phone camera after receipt of spoofed signals. Together Fig. \ref{25} and Fig. \ref{26} showing outdoor dyanmic spoofing alongwith movement of the GNSS receiver.}
  \label{26}
\end{figure}

Under outdoor conditions, the receiver was receiving genuine live sky signals. The Tx of spoofed GPS signals led to the non-reception of genuine GNSS signals by the receiver, as the generated signal strength was much higher than the genuine GNSS signal strength. It was found that the time taken by the GNSS receiver to lock onto the received spoofed signals increased under outdoor conditions. Also, the time to lock onto the received spoofed signals further increased when dynamic spoofing was carried out, especially when the GNSS receiver had achieved lock onto genuine signals before the receipt of spoofed signals. Fig. \ref{24} shows screenshots of Google Maps and OLA App before and after Tx of the spoofed signals under outdoor conditions. Figs. \ref{25} and  \ref{26} show the results during dynamic spoofing attacks being tested with the movement of the GNSS receiver, i.e., the Android mobile device. The movement of the Android mobile phone was in a straight line without any change in height, starting from a distance and ending at the spoofing signal generation system. This movement was imparted by walking with the Android mobile phone in hand.


\begin{table*}
  \caption{Results of various tests on different GNSS receivers.}
  \label{tab:table6}
  \begin{tabular}{c|cccc|cccc}
    \toprule
   \textbf{Location}&\multicolumn{4}{c|}{\textbf{Indoor}}&\multicolumn{4}{c}{\textbf{Outdoor}}\\
   \hline
\textbf{Spoofed Signal}&\textbf{S1$^{\mathrm{*}}$}&\textbf{S2$^{\mathrm{*}}$}&\textbf{S3$^{\mathrm{*}}$}& \textbf{S4$^{\mathrm{*}}$}&\textbf{S1$^{\mathrm{*}}$}&\textbf{S2$^{\mathrm{*}}$}&\textbf{S3$^{\mathrm{*}}$}& \textbf{S4$^{\mathrm{*}}$}\\
   \midrule
    {Static Spoofing}& \textcolor{blue}{S} & \textcolor{blue}{S}& \textcolor{red}{@}& \textcolor{blue}{S}& \textcolor{blue}{S} & \textcolor{blue}{S}& \textcolor{red}{@}& \textcolor{blue}{S}\\
    {Dynamic Spoofing}& \textcolor{blue}{S} & \textcolor{blue}{S}& \textcolor{red}{@}& \textcolor{blue}{S}& \textcolor{blue}{S} & \textcolor{blue}{S}& \textcolor{red}{@}& \textcolor{red}{@}\\
    \bottomrule
    \multicolumn{9}{p{.60\textwidth}}{$^{\mathrm{*}}$ Sr. No. as given in Table \ref{tab:table4}, \textcolor{blue}{S}= Successfully Spoofed,  \textcolor{red}{@}= did not get spoofed}
    \end{tabular}
\end{table*}

\subsection{Results of Static and Dynamic Spoofing under Indoor and Outdoor Conditions on Different GNSS Receivers}
\begin{figure}[h]
  \centering
  \includegraphics[width=\linewidth]{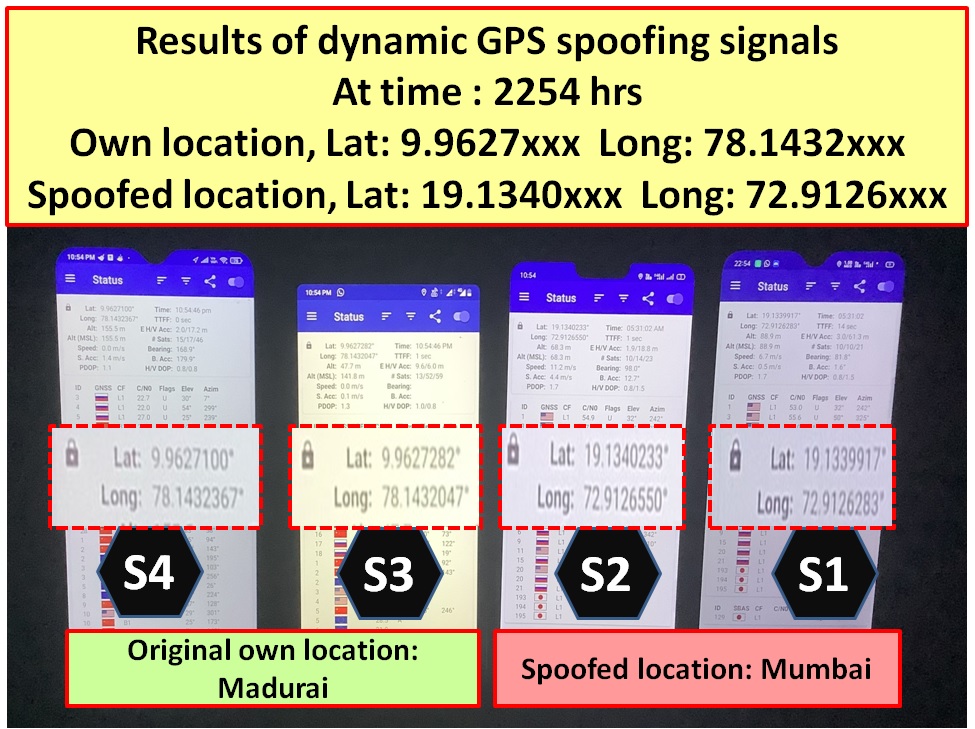}
  \caption{Results of dynamic GPS spoofing on different GNSS receivers.}
  \Description{Picture taken during receipt of dynamic GPS spoofed signals by multiple Android mobile phones based GNSS receivers.}
  \label{28}
\end{figure}

\begin{figure}[h]
  \centering
  \includegraphics[width=\linewidth]{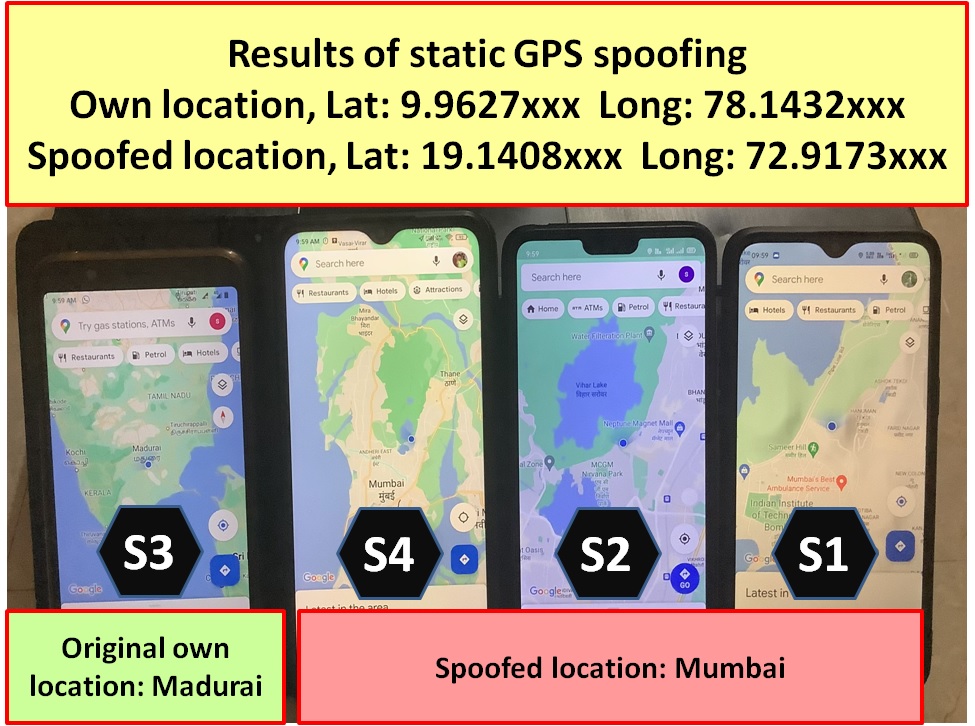}
  \caption{Results of static GPS spoofing on different GNSS receivers.}
  \Description{Picture taken during receipt of static GPS spoofed signals by multiple Android mobile phones based GNSS receivers.}
  \label{29}
\end{figure}

Table \ref{tab:table6} summarizes all the testing results. Fig. \ref{28} and Fig. \ref{29} show the results. The results show that in presence of live sky signals from three constellations, the receiver did not get spoofed by dynamic signals. Thus, receivers with more constellation support are less prone to get spoofed by dynamic signals (provided the genuine signals are not jammed). In addition, even a three-constellation-supported GNSS receiver can get spoofed if  receiver is overwhelmed by the spoofing signal, as in the case of static spoofing (S4 getting spoofed during static spoofing). The reason for exemplary resilience shown by S3 of not accepting the spoofed GPS signals, filtering it out and receiving only genuine signals is anti-spoofing measures being employed in it. This was researched and it was found that its chipset supports Observed Time Difference Of Arrival (OTDOA) \cite{636}, a Long Term Evolution (LTE) 4G based positioning technique \cite{fischer2014observed}. From our results, it can be seen that S3 utilised its OTDOA positioning and timing to filter out spoofed signals. 

\subsection{Results of Indoor and Outdoor Spoofing Against a Commercial Drone (DJI Mavic 2 Pro)}
\begin{figure}[h]
  \centering
  \includegraphics[width=\linewidth]{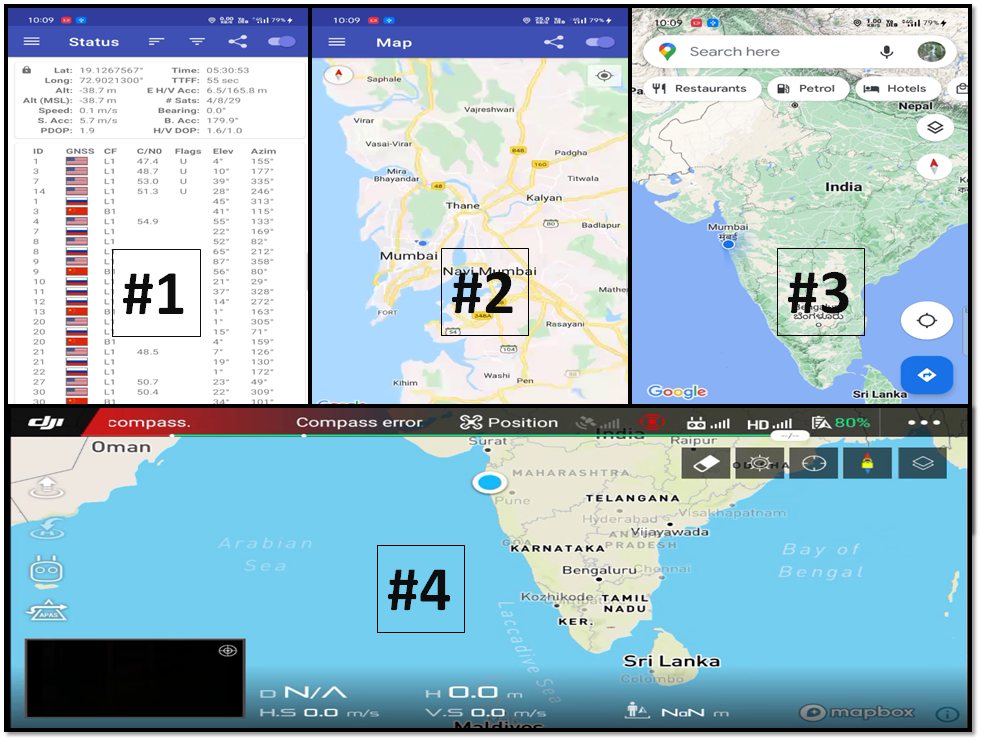}
  \caption{Screenshots taken on drone controller Android device during indoor spoofing: top left \#1: GPSTest App showing receipt of spoofed signals, top center \#2: GPSTest App map showing controller's location, top right \#3: Google Maps App showing controller's location, bottom \#4: DJI GO 4 App showing controller's location.}
  \Description{Picture showing results of indoor GPS spoofing attack on a commercial drones based on screenshots taken on drone controller Android device.}
  \label{DTI}
\end{figure}
\begin{figure}[h]
  \centering
  \includegraphics[width=\linewidth]{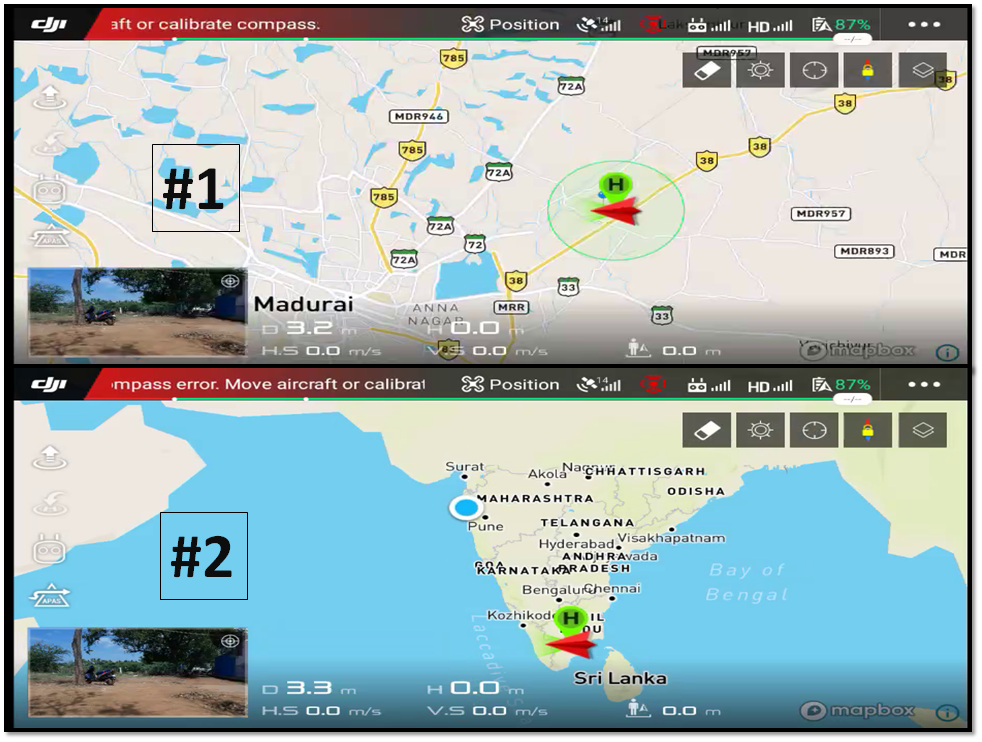}
  \caption{Screenshots taken on drone controller with DJI Go 4 App running. Top: drone own location; Bottom: controller's and drone's locations.}
  \Description{Picture showing results of outdoor GPS spoofing attack on a commercial drones based on screenshots taken on drone controller Android device.}
  \label{DTO}
\end{figure}
\begin{figure}[h]
  \centering
  \includegraphics[width=0.9\linewidth]{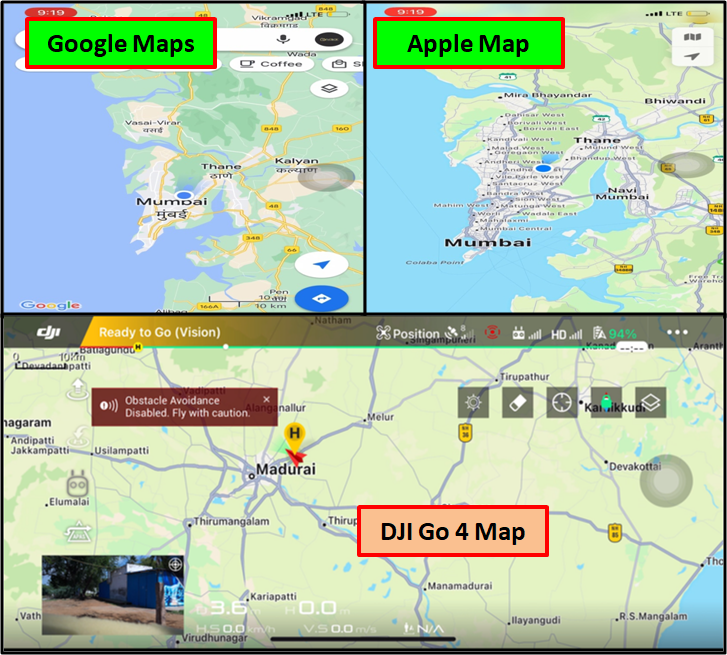}
  \caption{Screenshots taken on ios based drone controller with DJI Go 4 App running. Top left: Google Maps showing spoofed location; Top right: Apple map showing spoofed location: Bottom: DJI Go 4 map showing drone's location and registered home location.}
  \Description{Results received in an ios controller, which are similar to those received in an android based controller.}
  \label{DTOios}
\end{figure}
When tested indoor (in absence of genuine signals), the drone controller accepted the spoofed signals and showed the spoofed location as its own location. Our results of indoor spoofing against the drone can be seen in Fig. \ref{DTI}. When tested outdoor (once the drone and its controller had locked onto genuine GNSS signals), the drone controller accepted the spoofed signals and showed the spoofed location as its own location. However, the drone registered its original location as its home location (seen in the bottom half of Fig. \ref{DTO} as H circled by green color). The drone with its alignment is seen in the screenshots as a red arrow head. Fig. \ref{DTOios} shows similar results received in an iOS based drone controller. From the results received during indoor and outdoor testing against the drone, it was found that the drone did not accept the spoofed signals. It can be assumed that the manufacturers have incorporated countermeasures against GPS spoofing in their latest firmware update. Since we did not have proprietary permission, we could not find which particular countermeasure has been implemented in this drone. Our results show that a vulnerability which these drones  (DJI made) once had (shown in many research papers till late 2020) has been patched up. 

\section{Conclusions and Future Work}
\label{Conc}
Our results show that an asynchronous GPS spoofing attack is effective even under live sky signals against GNSS receivers in Android mobile phones with single frequency support. A DJI made commercial drone is found to be secure from asynchronous GPS spoofing. The obvious countermeasure against such a GPS spoofing attack is a multi-frequency GPS receiver or GNSS receiver. Since commercial paid GNSS simulators are capable of generating multi-frequency multi-constellation signals, a multi-frequency and a multi-constellation GNSS receiver can also be spoofed. Utilising simple anti-spoofing measures like filtering out high power signals or signals with time errors more than a few milliseconds can be sufficient to protect these high volume consumer device GNSS receivers from asynchronous GPS spoofing. A direction for future research is to explore the efficacy of GNSS simulator based spoofing attacks and countermeasures. It will also be interesting to see how countermeasures such as OTDOA (which utilizes the network for positioning) and those implemented in commercial drones will fare against GNSS simulator based synchronous spoofing attacks.
\begin{acks}
The work was partially supported by SEED grant [RD/0516-IRCCSH0-026 (16IRCCSG009)].
\end{acks}

\bibliographystyle{ACM-Reference-Format}
\bibliography{sample-base}


\begin{thebibliography}{40}


\ifx \showCODEN    \undefined \def \showCODEN     #1{\unskip}     \fi
\ifx \showDOI      \undefined \def \showDOI       #1{#1}\fi
\ifx \showISBNx    \undefined \def \showISBNx     #1{\unskip}     \fi
\ifx \showISBNxiii \undefined \def \showISBNxiii  #1{\unskip}     \fi
\ifx \showISSN     \undefined \def \showISSN      #1{\unskip}     \fi
\ifx \showLCCN     \undefined \def \showLCCN      #1{\unskip}     \fi
\ifx \shownote     \undefined \def \shownote      #1{#1}          \fi
\ifx \showarticletitle \undefined \def \showarticletitle #1{#1}   \fi
\ifx \showURL      \undefined \def \showURL       {\relax}        \fi
\providecommand\bibfield[2]{#2}
\providecommand\bibinfo[2]{#2}
\providecommand\natexlab[1]{#1}
\providecommand\showeprint[2][]{arXiv:#2}

\bibitem[Aabid et~al\mbox{.}(2022)]%
        {aabid2022reviews}
\bibfield{author}{\bibinfo{person}{Abdul Aabid}, \bibinfo{person}{Bisma
  Parveez}, \bibinfo{person}{Nagma Parveen}, \bibinfo{person}{Sher~Afghan
  Khan}, {and} \bibinfo{person}{Md Abdul}.} \bibinfo{year}{2022}\natexlab{}.
\newblock \showarticletitle{Reviews on Design and Development of Unmanned
  Aerial Vehicle (Drone) for Different Applications}.
\newblock \bibinfo{journal}{\emph{J. Mech. Eng. Res. Dev}}
  \bibinfo{volume}{45}, \bibinfo{number}{2} (\bibinfo{year}{2022}),
  \bibinfo{pages}{53--69}.
\newblock


\bibitem[Arteaga et~al\mbox{.}(2019)]%
        {arteaga2019analysis}
\bibfield{author}{\bibinfo{person}{Sandra~P{\'e}rez Arteaga},
  \bibinfo{person}{Luis Alberto~Mart{\'\i}nez Hern{\'a}ndez},
  \bibinfo{person}{Gabriel~S{\'a}nchez P{\'e}rez}, \bibinfo{person}{Ana
  Lucila~Sandoval Orozco}, {and} \bibinfo{person}{Luis Javier~Garc{\'\i}a
  Villalba}.} \bibinfo{year}{2019}\natexlab{}.
\newblock \showarticletitle{Analysis of the GPS spoofing vulnerability in the
  drone 3DR solo}.
\newblock \bibinfo{journal}{\emph{IEEE Access}}  \bibinfo{volume}{7}
  (\bibinfo{year}{2019}), \bibinfo{pages}{51782--51789}.
\newblock


\bibitem[Barbeau and Lockwood(2021)]%
        {GPSTest}
\bibfield{author}{\bibinfo{person}{Sean Barbeau} {and} \bibinfo{person}{Mike
  Lockwood}.} \bibinfo{year}{2021}\natexlab{}.
\newblock \bibinfo{booktitle}{\emph{{GPSTest android application}}}.
\newblock
\urldef\tempurl%
\url{https://github.com/barbeau/gpstest}
\showURL{%
\tempurl}


\bibitem[Blanch et~al\mbox{.}(2012)]%
        {blanch2012satellite}
\bibfield{author}{\bibinfo{person}{Juan Blanch}, \bibinfo{person}{Todd Walter},
  {and} \bibinfo{person}{Per Enge}.} \bibinfo{year}{2012}\natexlab{}.
\newblock \showarticletitle{Satellite Navigation for aviation in 2025}.
\newblock \bibinfo{journal}{\emph{Proc. IEEE}} \bibinfo{volume}{100},
  \bibinfo{number}{Special Centennial Issue} (\bibinfo{year}{2012}),
  \bibinfo{pages}{1821--1830}.
\newblock


\bibitem[Boyle(2020)]%
        {boyle2020drone}
\bibfield{author}{\bibinfo{person}{M.J. Boyle}.}
  \bibinfo{year}{2020}\natexlab{}.
\newblock \bibinfo{booktitle}{\emph{The Drone Age: How Drone Technology Will
  Change War and Peace}}.
\newblock \bibinfo{publisher}{Oxford University Press, Incorporated},
  \bibinfo{address}{UK}.
\newblock
\showISBNx{9780190635862}
\showLCCN{2019047301}
\urldef\tempurl%
\url{https://books.google.co.in/books?id=6wbcDwAAQBAJ}
\showURL{%
\tempurl}


\bibitem[cddis.nasa.gov(2022)]%
        {Brdc}
\bibfield{author}{\bibinfo{person}{cddis.nasa.gov}.}
  \bibinfo{year}{2022}\natexlab{}.
\newblock \bibinfo{title}{Broadcast Ephemeris DataFile}.
\newblock
\newblock
\urldef\tempurl%
\url{https://cddis.nasa.gov/Data$_$and$_$Derived$_$Products/GNSS/broadcast
  $_$ephemeris $_$data.html}
\showURL{%
\tempurl}


\bibitem[Dey et~al\mbox{.}(2018)]%
        {dey2018security}
\bibfield{author}{\bibinfo{person}{Vishal Dey}, \bibinfo{person}{Vikramkumar
  Pudi}, \bibinfo{person}{Anupam Chattopadhyay}, {and} \bibinfo{person}{Yuval
  Elovici}.} \bibinfo{year}{2018}\natexlab{}.
\newblock \showarticletitle{Security Vulnerabilities of Unmanned Aerial
  Vehicles and Countermeasures: An Experimental Study}. In
  \bibinfo{booktitle}{\emph{2018 31st International Conference on VLSI Design
  and 2018 17th International Conference on Embedded Systems (VLSID)}}.
  \bibinfo{publisher}{IEEE}, \bibinfo{address}{Pune, India},
  \bibinfo{pages}{398--403}.
\newblock
\urldef\tempurl%
\url{https://doi.org/10.1109/VLSID.2018.97}
\showDOI{\tempurl}


\bibitem[Ebinuma(2018)]%
        {GPS-SDR-SIM}
\bibfield{author}{\bibinfo{person}{Takuji Ebinuma}.}
  \bibinfo{year}{2018}\natexlab{}.
\newblock \bibinfo{booktitle}{\emph{{GPS-SDR-SIM}}}.
\newblock
\urldef\tempurl%
\url{https://github.com/osqzss/gps-sdr-sim}
\showURL{%
\tempurl}


\bibitem[EUSPA(2020)]%
        {EUSPA}
\bibfield{author}{\bibinfo{person}{EUSPA}.} \bibinfo{year}{2020}\natexlab{}.
\newblock \bibinfo{title}{GNSS User Technology Report 2020}.
\newblock
\newblock
\urldef\tempurl%
\url{https://www.euspa.europa.eu/european-space/euspace-market/gnss-market/gnss-user-technology-report}
\showURL{%
\tempurl}


\bibitem[EUSPA(2022)]%
        {EUSPAEO}
\bibfield{author}{\bibinfo{person}{EUSPA}.} \bibinfo{year}{2022}\natexlab{}.
\newblock \bibinfo{title}{EUSPA EO and GNSS Market report 2022}.
\newblock
\newblock
\urldef\tempurl%
\url{https://www.euspa.europa.eu/2022-market-report}
\showURL{%
\tempurl}


\bibitem[Fischer(2014)]%
        {fischer2014observed}
\bibfield{author}{\bibinfo{person}{Sven Fischer}.}
  \bibinfo{year}{2014}\natexlab{}.
\newblock \showarticletitle{Observed time difference of arrival (OTDOA)
  positioning in 3GPP LTE}.
\newblock \bibinfo{journal}{\emph{Qualcomm White Pap}} \bibinfo{volume}{1},
  \bibinfo{number}{1} (\bibinfo{year}{2014}), \bibinfo{pages}{1--62}.
\newblock


\bibitem[Gordon et~al\mbox{.}(2019)]%
        {gordon2019security}
\bibfield{author}{\bibinfo{person}{Joshua Gordon}, \bibinfo{person}{Victoria
  Kraj}, \bibinfo{person}{Ji~Hun Hwang}, {and} \bibinfo{person}{Ashok Raja}.}
  \bibinfo{year}{2019}\natexlab{}.
\newblock \showarticletitle{A security assessment for consumer wifi drones}. In
  \bibinfo{booktitle}{\emph{2019 IEEE International Conference on Industrial
  Internet (ICII)}}. IEEE, \bibinfo{publisher}{IEEE},
  \bibinfo{address}{Orlando, FL, USA}, \bibinfo{pages}{1--5}.
\newblock


\bibitem[Hartmann and Steup(2013)]%
        {hartmann2013vulnerability}
\bibfield{author}{\bibinfo{person}{Kim Hartmann} {and}
  \bibinfo{person}{Christoph Steup}.} \bibinfo{year}{2013}\natexlab{}.
\newblock \showarticletitle{The vulnerability of UAVs to cyber attacks-An
  approach to the risk assessment}. In \bibinfo{booktitle}{\emph{2013 5th
  international conference on cyber conflict (CYCON 2013)}}. IEEE,
  \bibinfo{publisher}{IEEE}, \bibinfo{address}{Tallinn, Estonia},
  \bibinfo{pages}{1--23}.
\newblock


\bibitem[Hathaway(2014)]%
        {hathaway2014melissa}
\bibfield{author}{\bibinfo{person}{ME Hathaway}.}
  \bibinfo{year}{2014}\natexlab{}.
\newblock \showarticletitle{MELISSA E. HATHAWAYa and JOHNN. STEWARTb aCouncil
  of Experts, Global Cyber Security Center (GCSEC)}.
\newblock \bibinfo{journal}{\emph{Best Practices in Computer Network Defense:
  Incident Detection and Response}}  \bibinfo{volume}{35}
  (\bibinfo{year}{2014}), \bibinfo{pages}{130}.
\newblock


\bibitem[He et~al\mbox{.}(2018)]%
        {he2018friendly}
\bibfield{author}{\bibinfo{person}{Daojing He}, \bibinfo{person}{Yinrong Qiao},
  \bibinfo{person}{Shiqing Chen}, \bibinfo{person}{Xiao Du},
  \bibinfo{person}{Wenjie Chen}, \bibinfo{person}{Sencun Zhu}, {and}
  \bibinfo{person}{Mohsen Guizani}.} \bibinfo{year}{2018}\natexlab{}.
\newblock \showarticletitle{A friendly and low-cost technique for capturing
  non-cooperative civilian unmanned aerial vehicles}.
\newblock \bibinfo{journal}{\emph{IEEE Network}} \bibinfo{volume}{33},
  \bibinfo{number}{2} (\bibinfo{year}{2018}), \bibinfo{pages}{146--151}.
\newblock


\bibitem[Horton and Ranganathan(2018)]%
        {horton2018development}
\bibfield{author}{\bibinfo{person}{Eric Horton} {and} \bibinfo{person}{Prakash
  Ranganathan}.} \bibinfo{year}{2018}\natexlab{}.
\newblock \showarticletitle{Development of a GPS spoofing apparatus to attack a
  DJI Matrice 100 Quadcopter}.
\newblock \bibinfo{journal}{\emph{The Journal of Global Positioning Systems}}
  \bibinfo{volume}{16}, \bibinfo{number}{1} (\bibinfo{year}{2018}),
  \bibinfo{pages}{1--11}.
\newblock


\bibitem[Huang and Yang(2015)]%
        {huang2015low}
\bibfield{author}{\bibinfo{person}{Ling Huang} {and} \bibinfo{person}{Qing
  Yang}.} \bibinfo{year}{2015}\natexlab{}.
\newblock \showarticletitle{Low-cost GPS simulator GPS spoofing by SDR}. In
  \bibinfo{booktitle}{\emph{Proc. DEFCON}}. \bibinfo{publisher}{DEFCON23},
  \bibinfo{address}{Las Vegas, Nevada}.
\newblock


\bibitem[Humphreys et~al\mbox{.}(2008)]%
        {humphreys2008assessing}
\bibfield{author}{\bibinfo{person}{Todd~E Humphreys}, \bibinfo{person}{Brent~M
  Ledvina}, \bibinfo{person}{Mark~L Psiaki}, \bibinfo{person}{Brady~W
  O'Hanlon}, \bibinfo{person}{Paul~M Kintner}, {et~al\mbox{.}}}
  \bibinfo{year}{2008}\natexlab{}.
\newblock \showarticletitle{Assessing the spoofing threat: Development of a
  portable GPS civilian spoofer}. In \bibinfo{booktitle}{\emph{Proceedings of
  the 21st International Technical Meeting of the Satellite Division of The
  Institute of Navigation (ION GNSS 2008)}}. \bibinfo{publisher}{Proceedings of
  the 21st International Technical Meeting of the Satellite Division of The
  Institute of Navigation (ION GNSS 2008)}, \bibinfo{address}{Savannah, GA},
  \bibinfo{pages}{2314--2325}.
\newblock


\bibitem[Joshi(2019)]%
        {article}
\bibfield{author}{\bibinfo{person}{Divya Joshi}.}
  \bibinfo{year}{2019}\natexlab{}.
\newblock \showarticletitle{Drone technology uses and applications for
  commercial, industrial and military drones in 2020 and the future}.
\newblock \bibinfo{journal}{\emph{Business Insider}}  \bibinfo{volume}{Dec}
  (\bibinfo{year}{2019}).
\newblock


\bibitem[Karunakaran(2022)]%
        {karunakaran2022swarm}
\bibfield{author}{\bibinfo{person}{A Karunakaran}.}
  \bibinfo{year}{2022}\natexlab{}.
\newblock \showarticletitle{Swarm Drones and Indian Academia}.
\newblock \bibinfo{journal}{\emph{Journal of Defence Studies}}
  \bibinfo{volume}{16}, \bibinfo{number}{1} (\bibinfo{year}{2022}),
  \bibinfo{pages}{73--81}.
\newblock


\bibitem[Khan et~al\mbox{.}(2021)]%
        {khan2021gps}
\bibfield{author}{\bibinfo{person}{Shah~Zahid Khan}, \bibinfo{person}{Mujahid
  Mohsin}, {and} \bibinfo{person}{Waseem Iqbal}.}
  \bibinfo{year}{2021}\natexlab{}.
\newblock \showarticletitle{On GPS spoofing of aerial platforms: a review of
  threats, challenges, methodologies, and future research directions}.
\newblock \bibinfo{journal}{\emph{PeerJ Computer Science}}  \bibinfo{volume}{7}
  (\bibinfo{year}{2021}), \bibinfo{pages}{e507}.
\newblock


\bibitem[Krishna and Murphy(2017)]%
        {krishna2017review}
\bibfield{author}{\bibinfo{person}{CG~Leela Krishna} {and}
  \bibinfo{person}{Robin~R Murphy}.} \bibinfo{year}{2017}\natexlab{}.
\newblock \showarticletitle{A review on cybersecurity vulnerabilities for
  unmanned aerial vehicles}. In \bibinfo{booktitle}{\emph{2017 IEEE
  International Symposium on Safety, Security and Rescue Robotics (SSRR)}}.
  IEEE, \bibinfo{publisher}{IEEE}, \bibinfo{address}{Shanghai, China},
  \bibinfo{pages}{194--199}.
\newblock


\bibitem[Kumar(2021)]%
        {kumar2021sky}
\bibfield{author}{\bibinfo{person}{Mamidala~Jagadesh Kumar}.}
  \bibinfo{year}{2021}\natexlab{}.
\newblock \bibinfo{title}{The sky is not the limit: The new rules give wings to
  the Drone Technology in India}.
\newblock , \bibinfo{numpages}{463--464}~pages.
\newblock


\bibitem[Labsat(2021)]%
        {satgen}
\bibfield{author}{\bibinfo{person}{Labsat}.} \bibinfo{year}{2021}\natexlab{}.
\newblock \bibinfo{booktitle}{\emph{{SatGen GNSS Signal Simulation Software}}}.
\newblock
\urldef\tempurl%
\url{https://www.labsat.co.uk/index.php/en/products/satgen-simulator-software}
\showURL{%
\tempurl}


\bibitem[Lisi(2020)]%
        {lisi2020gnss}
\bibfield{author}{\bibinfo{person}{Marco Lisi}.}
  \bibinfo{year}{2020}\natexlab{}.
\newblock \showarticletitle{GNSS User Technology Report 2020}.
\newblock \bibinfo{journal}{\emph{GEOmedia}} \bibinfo{volume}{24},
  \bibinfo{number}{5} (\bibinfo{year}{2020}).
\newblock


\bibitem[McNeff(2002)]%
        {mcneff2002global}
\bibfield{author}{\bibinfo{person}{Jules~G McNeff}.}
  \bibinfo{year}{2002}\natexlab{}.
\newblock \showarticletitle{The global positioning system}.
\newblock \bibinfo{journal}{\emph{IEEE Transactions on Microwave theory and
  techniques}} \bibinfo{volume}{50}, \bibinfo{number}{3}
  (\bibinfo{year}{2002}), \bibinfo{pages}{645--652}.
\newblock


\bibitem[Misra and Enge(2011)]%
        {misraglobal}
\bibfield{author}{\bibinfo{person}{P. Misra} {and} \bibinfo{person}{P. Enge}.}
  \bibinfo{year}{2011}\natexlab{}.
\newblock \bibinfo{booktitle}{\emph{Global Positioning System: Signals,
  Measurements, and Performance}}.
\newblock \bibinfo{publisher}{Ganga-Jamuna Press}, \bibinfo{address}{Lincoln,
  MA}.
\newblock
\showISBNx{9780970954428}
\urldef\tempurl%
\url{https://books.google.co.in/books?id=5WJOywAACAAJ}
\showURL{%
\tempurl}


\bibitem[Noh et~al\mbox{.}(2019)]%
        {noh2019tractor}
\bibfield{author}{\bibinfo{person}{Juhwan Noh}, \bibinfo{person}{Yujin Kwon},
  \bibinfo{person}{Yunmok Son}, \bibinfo{person}{Hocheol Shin},
  \bibinfo{person}{Dohyun Kim}, \bibinfo{person}{Jaeyeong Choi}, {and}
  \bibinfo{person}{Yongdae Kim}.} \bibinfo{year}{2019}\natexlab{}.
\newblock \showarticletitle{Tractor beam: Safe-hijacking of consumer drones
  with adaptive GPS spoofing}.
\newblock \bibinfo{journal}{\emph{ACM Transactions on Privacy and Security
  (TOPS)}} \bibinfo{volume}{22}, \bibinfo{number}{2} (\bibinfo{year}{2019}),
  \bibinfo{pages}{1--26}.
\newblock


\bibitem[Orolia(2021)]%
        {broadsim}
\bibfield{author}{\bibinfo{person}{Orolia}.} \bibinfo{year}{2021}\natexlab{}.
\newblock \bibinfo{title}{Skydel BroadSim GNSS Simulation Platform}.
\newblock
\newblock
\urldef\tempurl%
\url{https://www.orolia.com/document/broadsim-datasheet/}
\showURL{%
\tempurl}


\bibitem[Qualcomm.com(2022)]%
        {636}
\bibfield{author}{\bibinfo{person}{Qualcomm.com}.}
  \bibinfo{year}{2022}\natexlab{}.
\newblock \bibinfo{title}{Snapdragon 636 specifications}.
\newblock
\newblock
\urldef\tempurl%
\url{https://www.qualcomm.com/products/application/smartphones/snapdragon-6-series-mobile-platforms/snapdragon-636-mobile-platform-Overview}
\showURL{%
\tempurl}


\bibitem[Rustamov et~al\mbox{.}(2020)]%
        {rustamov2020assessment}
\bibfield{author}{\bibinfo{person}{Akmal Rustamov}, \bibinfo{person}{Neil
  Gogoi}, \bibinfo{person}{Alex Minetto}, {and} \bibinfo{person}{Fabio Dovis}.}
  \bibinfo{year}{2020}\natexlab{}.
\newblock \showarticletitle{Assessment of the vulnerability to spoofing attacks
  of gnss receivers integrated in consumer devices}. In
  \bibinfo{booktitle}{\emph{2020 international conference on localization and
  gnss (icl-gnss)}}. IEEE, \bibinfo{publisher}{IEEE},
  \bibinfo{address}{Tampere, Finland}, \bibinfo{pages}{1--6}.
\newblock


\bibitem[Saputro et~al\mbox{.}(2020)]%
        {saputro2020implementation}
\bibfield{author}{\bibinfo{person}{Jabang~Aru Saputro},
  \bibinfo{person}{Esa~Egistian Hartadi}, {and} \bibinfo{person}{Mohamad
  Syahral}.} \bibinfo{year}{2020}\natexlab{}.
\newblock \showarticletitle{Implementation of GPS Attacks on DJI Phantom 3
  Standard Drone as a Security Vulnerability Test}. In
  \bibinfo{booktitle}{\emph{2020 1st International Conference on Information
  Technology, Advanced Mechanical and Electrical Engineering (ICITAMEE)}}.
  IEEE, \bibinfo{publisher}{IEEE}, \bibinfo{address}{Yogyakarta, Indonesia},
  \bibinfo{pages}{95--100}.
\newblock


\bibitem[Sathyamoorthy et~al\mbox{.}(2020)]%
        {sathyamoorthy2020evaluation}
\bibfield{author}{\bibinfo{person}{Dinesh Sathyamoorthy}, \bibinfo{person}{Z
  Fitry}, \bibinfo{person}{Esa Selamat}, \bibinfo{person}{S Hassan},
  \bibinfo{person}{Ahmad Firdaus}, {and} \bibinfo{person}{Z Zaimy}.}
  \bibinfo{year}{2020}\natexlab{}.
\newblock \showarticletitle{Evaluation of the vulnerabilities of unmanned
  aerial vehicles (UAVs) to global positioning system (GPS) jamming and
  spoofing}.
\newblock \bibinfo{journal}{\emph{Defence S and T Technical Bulletin}}
  \bibinfo{volume}{13} (\bibinfo{year}{2020}), \bibinfo{pages}{333--343}.
\newblock


\bibitem[Shepard et~al\mbox{.}(2012)]%
        {shepard2012drone}
\bibfield{author}{\bibinfo{person}{Daniel~P Shepard},
  \bibinfo{person}{Jahshan~A Bhatti}, {and} \bibinfo{person}{Todd~E
  Humphreys}.} \bibinfo{year}{2012}\natexlab{}.
\newblock \showarticletitle{Drone hack: Spoofing attack demonstration on a
  civilian unmanned aerial vehicle.(2012)}.
\newblock \bibinfo{journal}{\emph{Google Scholar}} (\bibinfo{year}{2012}).
\newblock


\bibitem[Tippenhauer et~al\mbox{.}(2011)]%
        {tippenhauer2011requirements}
\bibfield{author}{\bibinfo{person}{Nils~Ole Tippenhauer},
  \bibinfo{person}{Christina P{\"o}pper}, \bibinfo{person}{Kasper~Bonne
  Rasmussen}, {and} \bibinfo{person}{Srdjan Capkun}.}
  \bibinfo{year}{2011}\natexlab{}.
\newblock \showarticletitle{On the requirements for successful GPS spoofing
  attacks}. In \bibinfo{booktitle}{\emph{Proceedings of the 18th ACM conference
  on Computer and communications security}}. \bibinfo{publisher}{Association
  for Computing Machinery}, \bibinfo{address}{New York, NY, United States},
  \bibinfo{pages}{75--86}.
\newblock


\bibitem[Tufekci and Tunc(2021)]%
        {tufekci2021vulnerability}
\bibfield{author}{\bibinfo{person}{Burak Tufekci} {and} \bibinfo{person}{Cihan
  Tunc}.} \bibinfo{year}{2021}\natexlab{}.
\newblock \showarticletitle{Vulnerability and Threat Analysis of UAVs}. In
  \bibinfo{booktitle}{\emph{2021 IEEE/ACS 18th International Conference on
  Computer Systems and Applications (AICCSA)}}. IEEE,
  \bibinfo{publisher}{IEEE}, \bibinfo{address}{Tangier, Morocco},
  \bibinfo{pages}{1--2}.
\newblock


\bibitem[{Wikipedia contributors}(2022)]%
        {GPSSignals}
\bibfield{author}{\bibinfo{person}{{Wikipedia contributors}}.}
  \bibinfo{year}{2022}\natexlab{}.
\newblock \bibinfo{title}{GPS signals --- {Wikipedia}{,} The Free
  Encyclopedia}.
\newblock
\newblock
\urldef\tempurl%
\url{https://en.wikipedia.org/w/index.php?title=GPS_signals&oldid=1091064831}
\showURL{%
\tempurl}
\newblock
\shownote{[Online; accessed 3-June-2022]}.


\bibitem[Yaacoub et~al\mbox{.}(2020)]%
        {yaacoub2020security}
\bibfield{author}{\bibinfo{person}{Jean-Paul Yaacoub}, \bibinfo{person}{Hassan
  Noura}, \bibinfo{person}{Ola Salman}, {and} \bibinfo{person}{Ali Chehab}.}
  \bibinfo{year}{2020}\natexlab{}.
\newblock \showarticletitle{Security analysis of drones systems: Attacks,
  limitations, and recommendations}.
\newblock \bibinfo{journal}{\emph{Internet of Things}}  \bibinfo{volume}{11}
  (\bibinfo{year}{2020}), \bibinfo{pages}{100218}.
\newblock


\bibitem[Zheng and Sun(2020)]%
        {zheng2020hijacking}
\bibfield{author}{\bibinfo{person}{Xian-Chun Zheng} {and}
  \bibinfo{person}{Hung-Min Sun}.} \bibinfo{year}{2020}\natexlab{}.
\newblock \showarticletitle{Hijacking unmanned aerial vehicle by exploiting
  civil GPS vulnerabilities using software-defined radio}.
\newblock \bibinfo{journal}{\emph{Sensors and Materials}} \bibinfo{volume}{32},
  \bibinfo{number}{8} (\bibinfo{year}{2020}), \bibinfo{pages}{2729--2743}.
\newblock


\bibitem[Zidan et~al\mbox{.}(2020)]%
        {zidan2020gnss}
\bibfield{author}{\bibinfo{person}{Jasmine Zidan}, \bibinfo{person}{Elijah~I
  Adegoke}, \bibinfo{person}{Erik Kampert}, \bibinfo{person}{Stewart~A
  Birrell}, \bibinfo{person}{Col~R Ford}, {and} \bibinfo{person}{Matthew~D
  Higgins}.} \bibinfo{year}{2020}\natexlab{}.
\newblock \showarticletitle{GNSS vulnerabilities and existing solutions: A
  review of the literature}.
\newblock \bibinfo{journal}{\emph{IEEE Access}}  \bibinfo{volume}{9}
  (\bibinfo{year}{2020}), \bibinfo{pages}{153960--153976}.
\newblock


\end{thebibliography}

\appendix

\end{document}